\documentclass[11pt]{article}

\usepackage[margin=1in]{geometry}
\usepackage{graphicx}
\usepackage{amsmath,amssymb,bm}
\usepackage{booktabs}
\usepackage{array}
\usepackage{tabularx}
\usepackage{microtype}
\usepackage[hidelinks]{hyperref}

\newcommand{\LLE}{\lambda_1}

\newcommand{\doi}[1]{\href{https://doi.org/#1}{doi:#1}}

\hypersetup{
  pdftitle={Equation-Free Period-Aware Forecast-Error Contraction for Estimating Negative Largest Lyapunov Exponents from Short Trajectory Ensembles},
  pdfauthor={Andrei Velichko, N'Gbo N'Gbo, Viet-Thanh Pham}
}
\title{\textbf{Equation-Free Period-Aware Forecast-Error Contraction for Estimating Negative Largest Lyapunov Exponents from Short Trajectory Ensembles}}

\author{
Andrei Velichko$^{1,*}$, N'Gbo N'Gbo$^{2,3}$, and Viet-Thanh Pham$^{4}$\\[0.75em]
\small $^{1}$Institute of Physics and Technology, Petrozavodsk State University,\\
\small 33 Lenina Prospekt, Petrozavodsk, Republic of Karelia, 185910, Russia\\
\small $^{2}$School of Science and Engineering, International University of Grand-Bassam,\\
\small Route de Bonoua, BP 564, Grand-Bassam, C\^ote d'Ivoire\\
\small $^{3}$Laboratory for Intelligence and Mathematics (LIMAs),\\
\small Route de Bonoua, BP 564, Grand-Bassam, C\^ote d'Ivoire\\
\small $^{4}$Faculty of Electronics Technology, Industrial University of Ho Chi Minh City,\\
\small Ho Chi Minh City, Vietnam\\[0.5em]
\small $^{*}$Corresponding author: \href{mailto:velichkogf@gmail.com}{velichkogf@gmail.com}
}
\date{August 2026}

\begin{document}
\maketitle

\begin{abstract}
Estimating positive largest Lyapunov exponents from data is comparatively natural because neighboring trajectories separate, whereas stable dynamics require resolving contraction before measurement noise or finite precision erases the signal. We introduce a period-aware forecast-error contraction procedure for estimating a dominant negative Lyapunov exponent from ensembles of short scalar trajectories without using governing equations or a Jacobian. A $k$-nearest-neighbor predictor is trained on trajectory histories, the geometric-mean absolute forecast error is evaluated at phase-consistent horizons, and the exponent is obtained from the slope of its logarithm. Unlike data-driven approaches that reconstruct local evolution matrices or differentiate a learned surrogate, the proposed method extracts the contraction rate directly from out-of-sample forecast errors. Two adaptations are essential: the forecast step is synchronized with the detected orbit period, and candidate slopes are accepted only when they form a stable consensus across several transient lengths. On the logistic map, the method recovers 92 of 112 negative-exponent parameter values with a mean absolute error of 0.0253 and $R^2=0.886$. On a two-dimensional map without fixed points, independent scalar pipelines based on the three observables $x_n$, $y_n$, and $z_n$ give mean absolute errors of 0.00879--0.01145 and $R^2=0.983$--0.986. Because the estimation stage uses only observed trajectories, the framework provides a basis for repeated-relaxation experiments in which short sensor responses are available but the governing equations and analytical Jacobian are unknown; experimental validation remains a subject of future work.
\end{abstract}


\paragraph{Significance statement.}
The largest Lyapunov exponent tells whether nearby states separate or converge. Data-driven estimators usually exploit separation and therefore naturally target positive exponents associated with chaos. Stable periodic dynamics pose a different problem: forecast errors contract, but the measurable contraction interval can be very short because trajectories rapidly become indistinguishable at machine precision. We show that negative exponents can nevertheless be recovered from short-trajectory ensembles when forecasts are compared at the same phase of a periodic orbit and when the fitted contraction rate is stable across several choices of transient length. The method uses the same forecast-error principle as our earlier positive-exponent estimator, but adds period synchronization and a consensus test tailored to stable regimes.

\section{Introduction}

Lyapunov exponents quantify the mean exponential rates at which infinitesimal perturbations evolve in a dynamical system. A positive largest Lyapunov exponent (LLE) is the standard signature of sensitive dependence, while a negative LLE indicates asymptotic contraction toward a stable fixed point or periodic orbit. Direct calculation from known equations is well established through tangent-map and orthogonalization algorithms.\cite{Benettin1980,Wolf1985} For measured scalar time series, classical approaches reconstruct local neighborhoods and track their separation.\cite{SanoSawada1985,Eckmann1986,Rosenstein1993,Kantz1994} These methods and their descendants have been used primarily in chaotic regimes, where the signal of interest grows above observational and numerical floors.

Our previous work introduced a machine-learning interpretation of the same idea: a predictor generates out-of-sample forecasts at several horizons, and the positive LLE is inferred from the exponential growth of the geometric-mean absolute forecast error.\cite{Velichko2025} The approach was validated on one-dimensional chaotic maps and was intentionally designed for positive exponents. In the present study, the same forecast-error principle is retained, but the data-collection stage is harmonized with the stable setting by generating ensembles of short post-transient trajectories. This ensemble-based reformulation is also used to recompute the positive logistic-map branch shown only as context in Fig.~\ref{fig:logistic}(a).

Direct model-free recovery of contracting Lyapunov directions is much less developed than positive-LLE estimation. A small group of studies has nevertheless demonstrated that negative exponents can be inferred from observations. Yang and Wu combined neighbor averaging with nonlinear local mappings to improve negative-exponent recovery from noisy time series, and a subsequent multidimensional formulation estimated all-negative and mixed-sign spectra.\cite{YangWu2011,YangWuZhang2012} Sun and Wu fitted a radial-basis-function surrogate to a scalar record and obtained Lyapunov spectra by differentiating the learned model.\cite{SunWu2012} More recent machine-learning work has estimated local or complete spectra from trajectory windows or single-variable series, although such methods commonly require supervised tangent-space labels or a trained surrogate and have focused mainly on chaotic dynamics.\cite{Ayers2022,MayoraCebollero2024} These studies establish that equation-free inference of negative exponents is possible, but they typically reconstruct a differentiable local dynamics and then apply matrix-based Lyapunov calculations.

For stable periodic motion, the related literature usually targets Floquet or characteristic multipliers rather than a negative LLE directly. Bias-corrected cycle-to-cycle regression has been developed for noisy rhythmic motion,\cite{AhnHogan2015} model-free Floquet multipliers have been extracted from scalar power-system measurements,\cite{Choi2019} and empirical characteristic-multiplier estimation has been improved through a change of basis that reduces noise amplification.\cite{Little2020} Data-driven Poincar\'e maps, phase--isostable reductions, and phase-amplitude reconstruction provide further routes to orbital stability and transverse decay rates.\cite{Bramburger2020,Wilson2020,Namura2022} In parallel, Koopman methods estimate decay eigenvalues, eigenfunctions, regions of attraction, or stability-relevant spectral structure from snapshot data, while data-driven contraction approaches learn or certify contracting behavior rather than a single dominant negative exponent.\cite{Williams2015,Mauroy2016,Matavalam2024,Han2025,Oliveira2025}

The data structure required by such an estimator also occurs naturally outside benchmark equations. Free-decay and ring-down measurements of bridges and other structures, event-wise damping analysis, and output-only modal monitoring all extract recovery information from sensor records, but they typically report logarithmic decrements, damping ratios, or stable modal poles rather than a negative Lyapunov exponent.\cite{Brownjohn2010,Lorenzoni2019,LopezAragon2024,Rosso2023} Short experimental trajectories are also increasingly used for equation-free inference in nonequilibrium systems and for nonlinear model discovery from video observations.\cite{Manikandan2021,Yang2024} Related vibration studies have estimated principal Lyapunov exponents from observed impact signals, while ensemble-based methods have inferred LLE expressions from a small number of random-state samples.\cite{Lee2024,ChenJinHuang2022} These developments motivate a direct scalar contraction estimator for repeated recovery responses, without claiming that a Lyapunov exponent is interchangeable with conventional modal damping.

In a stable regime, a direct sign reversal of a positive-LLE estimator is insufficient. Periodic trajectories alternate between distinct phases, so consecutive horizons need not follow one contraction envelope. The fitted slope also depends strongly on how much transient motion is removed, and strongly negative exponents drive errors to exact zeros or to a numerical floor before a sufficiently long profile can be observed. The present work therefore takes a deliberately narrower route than full-spectrum reconstruction: it estimates the dominant observable contraction rate directly from the decay of out-of-sample multi-horizon forecast errors, without reconstructing a Jacobian or differentiating a learned surrogate. To our knowledge, the reviewed literature has not combined short-trajectory ensembles, period-synchronized forecast horizons, and transient-slope consensus in this form.

This Fast Track study focuses on a single question: \emph{Can a forecast-error method estimate negative LLEs reliably from short trajectory ensembles?} Two benchmarks are used. The logistic map provides a one-dimensional system with directly computable reference exponents and narrow periodic windows. The second benchmark is a two-dimensional map with no fixed points, whose bifurcation sequence begins with a period-two cycle and proceeds toward chaos.\cite{Huynh2019} This system tests whether the contraction rate can be recovered from different scalar observations of a higher-dimensional state.

\section{Forecast-error contraction method}

\subsection{Short-trajectory ensemble and multi-horizon prediction}

Let $u_i(n)$ denote the scalar observation from trajectory $i=1,\ldots,N$. The trajectories are divided into training and test subsets. At a candidate transient length $T$, each trajectory is represented by the history vector
\begin{equation}
\bm{v}_i(T)=\big[u_i(T),u_i(T+\tau),\ldots,u_i(T+(Y-1)\tau)\big],
\label{eq:history}
\end{equation}
where $Y$ is the history length and $\tau$ is the history step. A $k$-nearest-neighbor predictor identifies $K$ training histories nearest to each test history and averages their future scalar values. For an actual forecast horizon $h$, the test error is
\begin{equation}
 e_i(h)=\left|u_i(T+(Y-1)\tau+h)-\widehat{u}_i(h)\right|.
\end{equation}
We aggregate across the test ensemble using the geometric mean
\begin{equation}
 E(h)=\exp\left[\frac{1}{N_{\rm test}}\sum_{i=1}^{N_{\rm test}}\log\big(\max[e_i(h),\epsilon]\big)\right],
\label{eq:gmae}
\end{equation}
which limits domination by a small number of large errors. In a locally contracting regime,
\begin{equation}
 \log E(h)\simeq C+\LLE h, \qquad \LLE<0,
\label{eq:slope}
\end{equation}
so the fitted slope estimates the largest negative exponent.

The computational pipeline is identical whether the short trajectories are obtained from repeated realizations, controlled perturbations, or segmentation of a sufficiently variable record. The distinction matters physically: segmentation of a fully settled periodic orbit does not recreate transverse perturbations. Thus, negative-LLE estimation requires either repeated realizations, naturally occurring perturbations, or transient segments that retain measurable spread. For the positive logistic-map context curve in Fig.~\ref{fig:logistic}(a), we use exactly the same ensemble-generation protocol but fit a growing profile after a fixed transient $T=1000$ with $Y=1$ and $K=3$. This comparison is useful because it isolates the effect of the sign of the exponent while keeping the sampling strategy compatible with the negative-LLE experiments.

\subsection{Period-aware horizons}

For a stable orbit of period $p$, horizons $1,2,3,\ldots$ compare different phases and can produce an oscillatory error profile. We estimate the period through the recurrence statistic
\begin{equation}
 Q_p(T)=\mathop{\rm median}_{i,n}\left|u_i(n+p)-u_i(n)\right|,
\label{eq:period}
\end{equation}
computed over a short window after $T$ for $p=1,\ldots,p_{\max}$. The smallest period close to the minimum of $Q_p$ is selected and confirmed in a shifted window. Candidate profiles are then evaluated at
\begin{equation}
 h_m=m s,\qquad s\in\{1,p\},\qquad m=1,\ldots,H.
\end{equation}
The choice $s=p$ compares the same phase of the periodic orbit and was preferred in most accepted stable cases.

The same scalar period statistic in Eq.~\eqref{eq:period} is applied independently to every observable. For the two-dimensional benchmark, the complete pipeline---period detection, history construction, nearest-neighbor prediction, forecast-error evaluation, parameter selection, and transient-slope consensus---is therefore repeated separately for $u_n=x_n$, $u_n=y_n$, and $u_n=\sqrt{x_n^2+y_n^2}$. No coordinates are combined during either period detection or exponent estimation. This identical scalar treatment makes the agreement among the three estimates a direct test of sensitivity to the observation function.

\subsection{Stable-transient consensus}

A high linear-fit $R^2$ alone does not guarantee the asymptotic contraction rate: an early transient can be almost perfectly linear but have the wrong slope. We therefore evaluate several candidate transient lengths $T_j$. At each $T_j$, accepted profiles must have a negative slope, sufficient linearity, predominantly decreasing log-error values, and limited contamination by exact-zero or numerical-floor errors. The candidate slopes are grouped across neighboring transient lengths. The final estimate is the median of a stable group,
\begin{equation}
 \widehat{\LLE}=\mathop{\rm median}_{j\in\mathcal{C}}\widehat{\LLE}(T_j),
\label{eq:consensus}
\end{equation}
where $\mathcal{C}$ contains mutually consistent slopes. A group of at least three transient lengths is labeled \emph{reliable}, a consistent pair is \emph{acceptable}, and an isolated slope is rejected. This last rule is essential in narrow periodic windows, where a single visually convincing profile can be strongly biased.

\paragraph*{Experimental-use protocol.}
A practical implementation would (i) collect short responses following comparable natural or controlled perturbations; (ii) align complete records at the perturbation onset or at a common oscillation phase; (iii) divide whole realizations, rather than individual samples, into training and test sets; (iv) detect a dominant recurrence period when the attracting response is periodic; (v) evaluate phase-consistent multi-horizon forecast errors; and (vi) repeat the slope fit over several transient offsets and report the median consensus together with its reliability class. The realizations may come from repeated experiments, event-triggered ring-down segments, or comparable operating cycles. No governing equations or analytical Jacobian are used during this estimation stage.

\subsection{Benchmarks and reference exponents}

The one-dimensional benchmark is the logistic map
\begin{equation}
 x_{n+1}=r x_n(1-x_n).
\label{eq:logistic}
\end{equation}
Reference exponents are calculated from
\begin{equation}
 \LLE(r)=\lim_{M\to\infty}\frac{1}{M}\sum_{n=1}^{M}
 \log\left|r(1-2x_n)\right|.
\label{eq:logref}
\end{equation}
The main sweep uses $r\in[3.5,4.0]$, $N=5000$ trajectories, a 70/30 training-test split, $K=3$, and $Y=1$. Candidate periods up to 16 and profile lengths $H=5,\ldots,10$ are tested. The previously published positive-LLE estimator is shown only to demonstrate continuity over the complete parameter interval; all new analysis is restricted to $\LLE<0$.

The two-dimensional benchmark is\cite{Huynh2019}
\begin{subequations}
\begin{align}
 x_{n+1}&=x_n+y_n,\\
 y_{n+1}&=y_n-a|y_n|-x_ny_n+b x_n^2-cy_n^2+d,
\end{align}
\label{eq:nofixed}
\end{subequations}
with $a=0.01$, $b=d=0.1$, and $c\in[1.7,2.0]$. For positive $b$ and $d$, Eq.~(\ref{eq:nofixed}) has no real fixed point; its bifurcation sequence therefore begins with a period-two orbit. Reference exponents are obtained from QR accumulation of the Jacobian after 5000 transient steps and over 10,000 subsequent iterations. The negative region contains 948 sampled values of $c$. Forecast-error estimates use $N=5000$ trajectories and the scalar observations $x_n$, $y_n$, and $z_n=\sqrt{x_n^2+y_n^2}$.

\section{Results}

\subsection{A transparent fixed-point example}

\begin{figure}[t]
\includegraphics[width=0.98\textwidth]{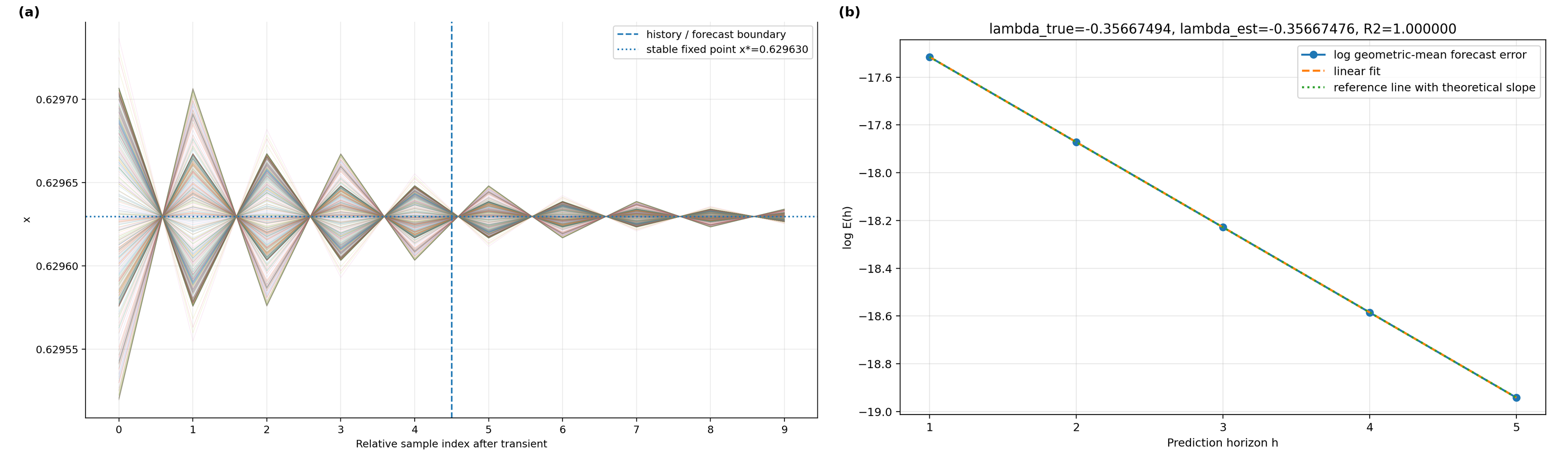}
\caption{Forecast-error contraction for a representative stable logistic-map point, $r=2.7$. (a) Ensemble of 500 trajectories after a transient of $T=20$; the dashed vertical line separates the five-sample history and the forecast interval. (b) Geometric-mean forecast error in logarithmic coordinates. The fitted slope and theoretical negative LLE are visually indistinguishable.}
\label{fig:demo}
\end{figure}

Figure~\ref{fig:demo} illustrates the contraction mechanism for the logistic map at $r=2.7$. The stable fixed point is $x^*=1-1/r=0.62963$, and the theoretical exponent is
\begin{equation}
 \LLE=\log|2-r|=\log(0.7)=-0.35667494.
\end{equation}
An ensemble of 500 trajectories is evaluated after $T=20$ with $Y=5$, $K=3$, and $H=5$. The alternating contraction toward the fixed point is directly visible in Fig.~\ref{fig:demo}(a). The five-point log-error profile in Fig.~\ref{fig:demo}(b) gives $\widehat{\LLE}=-0.35667476$, an absolute deviation of $1.8\times10^{-7}$, with $R^2$ indistinguishable from unity. This example is intentionally simple; it demonstrates what the method measures before the more difficult parameter sweeps are considered.

\subsection{Logistic-map parameter sweep}

\begin{figure}[t]
\includegraphics[width=0.98\textwidth]{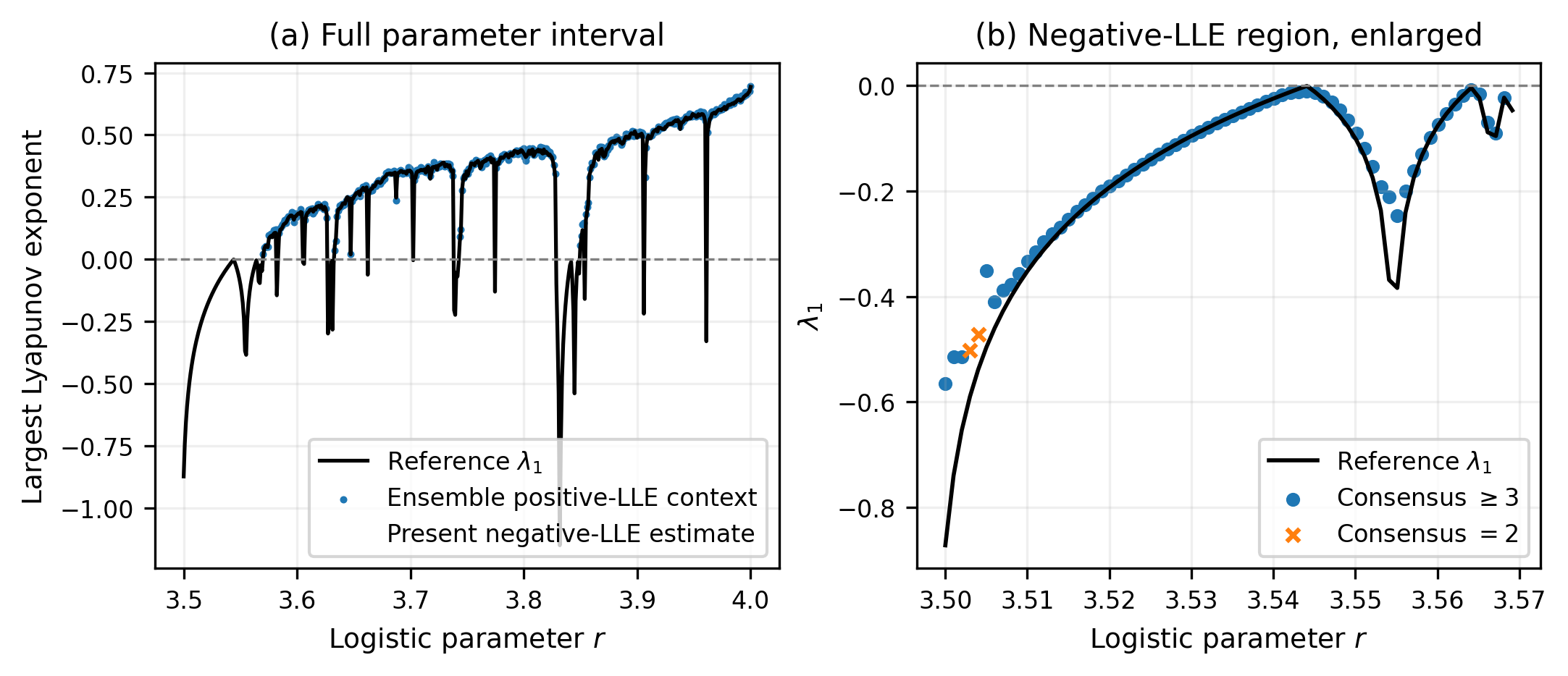}
\caption{Logistic-map results. (a) Reference LLE over $3.5\le r\le4.0$, an ensemble-based positive-branch context curve obtained with $T=1000$, $Y=1$, $K=3$, and present negative-LLE estimates. The positive results are included only as context; the contribution of this work is the negative branch. (b) Enlarged initial negative-LLE interval. Estimates supported by at least three transient lengths are shown separately from two-transient consensus estimates.}
\label{fig:logistic}
\end{figure}

Figure~\ref{fig:logistic}(a) places the new negative-LLE estimates in the context of the positive regime. The reference curve contains broad chaotic intervals separated by narrow periodic windows. The positive branch is shown using an ensemble-based reformulation of the earlier forecast-error-growth estimator, while the present contraction method populates the stable branch. The two procedures therefore use the same observable quantity, $\log E(h)$, but extract opposite slopes under different profile-selection rules.

For the 112 negative reference values, six isolated single-transient fits were rejected because they lacked a genuine consensus. The retained set contains 92 points (82.14\% coverage), with mean absolute error (MAE) 0.0253, root-mean-square error (RMSE) 0.0586, median absolute error 0.00488, and $R^2=0.886$. Most accepted points are much more accurate than the RMSE suggests; the average is dominated by several deep, narrow stable windows and by the strongly contracting beginning of the interval. Figure~\ref{fig:logistic}(b) enlarges $3.50\le r\le3.57$. The method closely follows the broad approach to the accumulation point, but underestimates the magnitude of the most negative exponents because the ensemble collapses before a long asymptotic profile can be measured. The full-interval panel in Fig.~\ref{fig:logistic}(a) also shows an ensemble-based positive-branch context calculation produced with the same short-trajectory generation strategy. For that branch, the fixed setting $T=1000$, $Y=1$, $K=3$, $s=1$, and $H=5$ yields MAE $=0.00709$ and $R^2=0.9964$, demonstrating that the earlier forecast-error-growth methodology remains effective after the sampling procedure is adapted from a single long trajectory to an ensemble of short post-transient realizations.

\subsection{Two-dimensional map without fixed points}

\begin{figure}[t]
\includegraphics[width=0.98\textwidth]{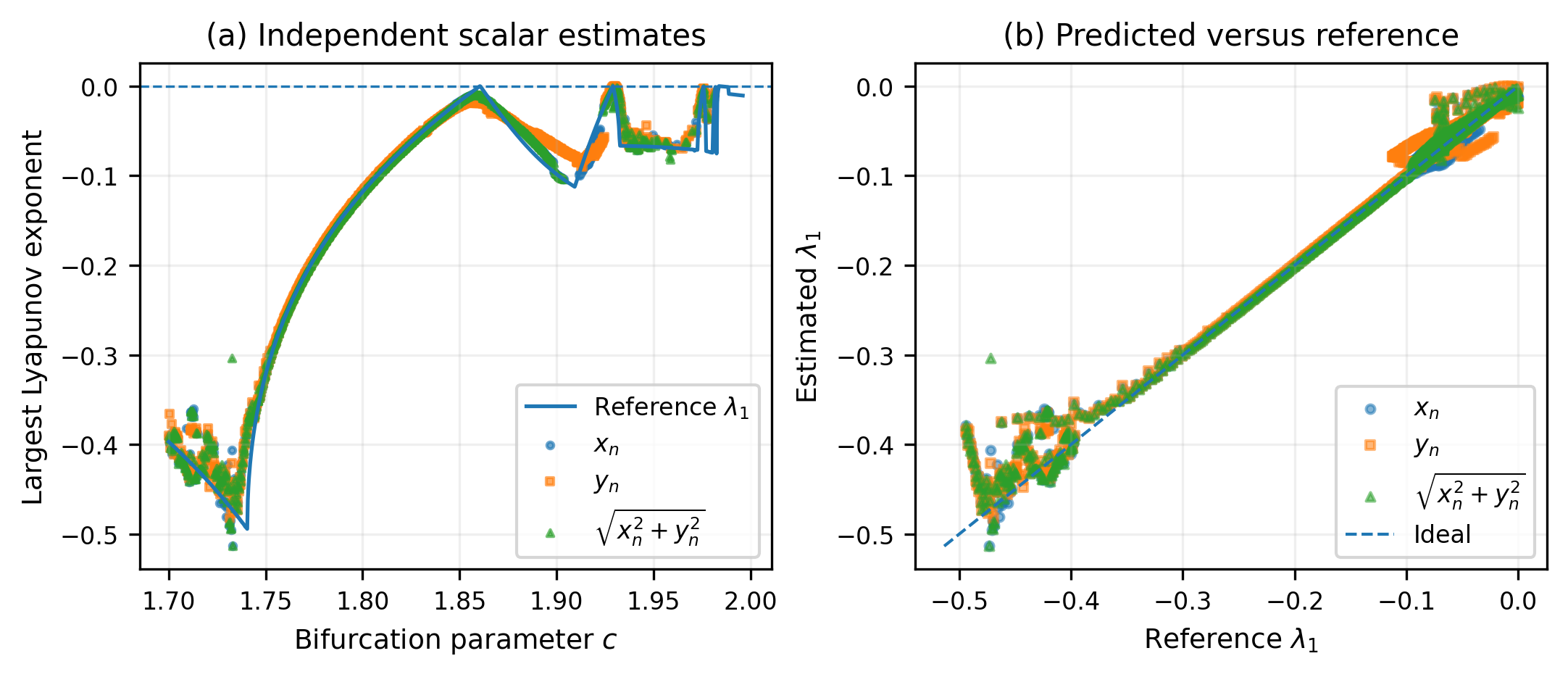}
\caption{Negative largest Lyapunov exponent for the two-dimensional map without fixed points. (a) Reference exponent and estimates obtained by applying the complete scalar pipeline independently to $x_n$, $y_n$, and $\sqrt{x_n^2+y_n^2}$. (b) Predicted-versus-reference representation. The dashed diagonal indicates exact agreement.}
\label{fig:2d}
\end{figure}

The no-fixed-point map provides a more demanding validation because the target exponent belongs to a two-dimensional state, while the prediction error is formed from a single scalar observation. Figure~\ref{fig:2d}(a) shows that all three independently processed observations reproduce the main reference curve throughout the negative-LLE interval. MAE is 0.00909 for $x_n$, 0.01145 for $y_n$, and 0.00879 for $\sqrt{x_n^2+y_n^2}$. Their corresponding $R^2$ values are 0.9858, 0.9830, and 0.9852. Coverage ranges from 83.23\% to 90.40\%.

The scatter plot in Fig.~\ref{fig:2d}(b) confirms close agreement for the three scalar observation functions. The coordinate $y_n$ gives the largest coverage, $x_n$ gives the smallest RMSE and the largest $R^2$, and the state norm gives the smallest median error. Because period detection and exponent estimation were both performed independently for each one-dimensional series, this agreement was obtained without using full-state information. The main residual bias is positive, meaning that the estimated exponents tend to be slightly less negative than the reference values.

\begin{table}[t]
\caption{Summary of forecast-error LLE estimation. The positive logistic-map row is prior-work context; the remaining rows are the negative-LLE results of this study. Logistic negative metrics require a transient consensus count of at least two.}
\label{tab:metrics}
\centering
\small
\setlength{\tabcolsep}{3.5pt}
\begin{tabularx}{\textwidth}{>{\raggedright\arraybackslash}Xrrrrrr}
Benchmark and observable & Accepted/total & Coverage (\%) & MAE & RMSE & Median AE & $R^2$\\
\hline
Logistic, positive $x_n$ (ensemble context; $T=1000$, $Y=1$) & 388/388 & 100.00 & 0.00709 & 0.00900 & 0.00599 & 0.9964\\
Logistic, negative $x_n$ & 92/112 & 82.14 & 0.02527 & 0.05863 & 0.00488 & 0.8863\\
No-fixed-point map, $x_n$ & 832/948 & 87.76 & 0.00909 & 0.01812 & 0.00227 & 0.9858\\
No-fixed-point map, $y_n$ & 857/948 & 90.40 & 0.01145 & 0.01956 & 0.00488 & 0.9830\\
No-fixed-point map, $\sqrt{x_n^2+y_n^2}$ & 789/948 & 83.23 & 0.00879 & 0.01882 & 0.00133 & 0.9852\\
\end{tabularx}
\end{table}

\section{Discussion}

The experiments identify two adaptations that separate successful negative-LLE estimation from a naive fit of decreasing prediction errors. First, the horizon step should respect the orbit period. In a period-$p$ regime, using $s=p$ compares the same orbit phase and converts an alternating or multibranch profile into a single contraction envelope. Second, the transient length cannot be selected by fit linearity alone. Early transients can produce $R^2>0.99$ while their slopes differ substantially from the asymptotic exponent. Requiring agreement across several transient lengths reduces this failure mode without using the reference LLE during selection.

The proposed estimator is not the first data-driven method capable of recovering negative Lyapunov exponents. Earlier nonlinear-mapping and RBF-network approaches reconstruct a local differentiable dynamics from a time series and then apply matrix-based Lyapunov-spectrum calculations.\cite{YangWu2011,YangWuZhang2012,SunWu2012} Their main advantage is the possibility of estimating several exponents, including contracting directions, and some formulations explicitly address additive noise. Their cost is the need to select a phase-space reconstruction, identify a sufficiently accurate local map, differentiate or linearize that map, and propagate a sequence of estimated Jacobians. Our objective is more restricted: only the dominant observable contraction rate is sought. This restriction avoids local Jacobian reconstruction and permits a direct scalar estimator based on out-of-sample forecast errors.

The period-aware horizon can be interpreted as a scalar counterpart of the one-period comparison underlying empirical Floquet analysis.\cite{AhnHogan2015,Choi2019,Little2020} The important difference is that no transition matrix is fitted: the decay rate is extracted from the prediction-error envelope. Likewise, Koopman and phase--isostable methods can recover decay spectra or transverse coordinates from data,\cite{Williams2015,Mauroy2016,Wilson2020,Namura2022} whereas the present procedure targets one directly measurable slope and supplies an explicit reliability decision. The stable-transient consensus addresses a failure mode that is especially visible for strongly negative exponents: a profile may be highly linear while representing an early transient rather than the asymptotic contraction rate.

The ensemble-based positive logistic-map context calculation also helps position the present work relative to our previous positive-LLE study. In Ref.~\cite{Velichko2025}, the logistic-map KNN estimator reached $R^2_{\mathrm{pos}}=0.998$ on the positive branch. The ensemble-based reformulation used here gives $R^2=0.9964$, supporting the practical compatibility of long-record and repeated-short-trajectory sampling in the chaotic regime. Thus, the small positive-branch experiment is not a second focus of the paper, but it shows that the short-trajectory protocol introduced for stable dynamics preserves the performance of the earlier forecast-error-growth formulation.

The logistic and two-dimensional results differ in a physically informative way. The no-fixed-point map contains extended smooth parameter branches with period doubling, and the stable-transient consensus tracks them with $R^2=0.983$--0.986 across three independently processed scalar observations. The logistic map contains extremely narrow periodic windows and superstable neighborhoods in which the true exponent changes sharply. There, a finite trajectory ensemble may pass directly from an incompletely settled transient to a numerical floor. This produces missing estimates or a conservative bias toward zero. Such cases are not merely statistical outliers; they expose an identifiability limit of contraction-based analysis at finite precision.

The ensemble formulation provides a natural route toward experimental use because each trajectory may represent a short recovery response following a comparable disturbance rather than a numerically generated initial condition. Structural free-decay and bridge ring-down records are especially transparent examples: current practice estimates damping from envelope decay or output-only modal poles, and reported damping values may depend strongly on filtering, mode separation, and the identification procedure.\cite{Lorenzoni2019,LopezAragon2024,Rosso2023} The proposed estimator would instead compare complete realizations through out-of-sample forecast contraction. It is not a replacement for modal analysis and does not imply that an LLE equals a damping ratio; it supplies a nonlinear, observation-dependent return rate that can be compared with established decay measures on the same events.

The physical interpretation depends on the attractor. For convergence toward a stable equilibrium or for a discrete-time or Poincar\'e map, the slope may approximate the dominant negative LLE. For an autonomous continuous-time limit cycle, the largest exponent is zero because of phase neutrality; a phase-aligned or stroboscopic implementation instead targets a transverse Lyapunov or Floquet decay exponent. The broader term \emph{dominant observable contraction rate} is therefore preferable when only a scalar experimental response is available.

The present method should therefore be interpreted as an estimator of a measurable contraction interval, not as a replacement for tangent-space calculation when equations are known. Its practical value is greatest when repeated short realizations are available but a reliable mechanistic model or Jacobian is not. Koopman and extended-DMD approaches provide a complementary equation-free alternative by estimating decay eigenvalues in a chosen observable space, with increasingly rigorous spectral and finite-data formulations.\cite{Korda2018,Brunton2022,Colbrook2024} Our method is narrower: it avoids selecting a global observable dictionary and returns one forecast-based slope with an explicit accept/reject decision. In this setting, rejecting a parameter point is preferable to reporting a highly precise linear fit with an unstable slope; the reliable/acceptable/unreliable classification is consequently part of the estimator rather than a post-processing label.

The earlier positive-exponent forecast-error formulation retained useful accuracy under moderate additive noise, indicating that the predictive principle itself is not intrinsically restricted to noise-free data.\cite{Velichko2025} That result cannot be transferred automatically to the present contraction regime: negative exponents drive the forecast error toward the observational noise floor, so the most strongly contracting cases may become unidentifiable earlier. A dedicated noise study must therefore quantify recoverable exponent magnitude, coverage, and the behavior of the transient-consensus rule.

Several limitations remain. Measurement noise introduces an error floor earlier than floating-point arithmetic and will likely reduce the recoverable magnitude of negative exponents. The ensemble size, perturbation amplitude, and predictor neighborhood determine how long contraction remains observable. Finally, the algorithm has been tested on deterministic maps but not yet on experimental oscillators or continuous-time systems.

A like-for-like numerical comparison with nonlinear local-mapping, surrogate-Jacobian, empirical Floquet, and Koopman estimators remains future work because these approaches require different data structures and target partly different stability quantities.

\section{Conclusion}

We extended forecast-error Lyapunov analysis from divergence to contraction. The method estimates a dominant negative exponent by fitting the decay of geometric-mean multi-horizon prediction errors, synchronizing forecast horizons with the detected orbit period, and requiring slope stability across several transient lengths. Unlike local-mapping and surrogate-Jacobian approaches, it does not reconstruct a differentiable dynamical model or propagate estimated tangent matrices. The logistic-map test demonstrates both the capability and the finite-precision limit of the approach: most accepted points have small errors, while strongly stable and superstable neighborhoods remain difficult. The two-dimensional no-fixed-point map gives substantially higher global agreement and shows that the inferred contraction rate is nearly invariant across three scalar observations. The ensemble-based positive logistic-map calculation also remains close to the performance reported in our earlier positive-LLE study, showing that one short-trajectory protocol can support both forecast-error growth and contraction analyses.

The method is therefore best viewed as an equation-free estimator of the dominant observable contraction rate from repeated short relaxation trajectories. Depending on the dynamical setting, this rate may correspond to a negative largest Lyapunov exponent, a transverse exponent, or a Floquet decay rate. Although the present validation uses maps with known reference exponents, the estimation procedure itself requires only complete short realizations and is directly compatible with event-triggered recovery records from sensors. This opens a path toward structural ring-down, electromechanical recovery, controlled periodic motion, and other repeated-relaxation experiments, while leaving experimental validation and comparison with conventional damping or modal-pole estimates for future work. Further work should quantify robustness to measurement noise, trajectory count, perturbation amplitude, and partial observation, and validate the method on repeated experimental relaxation records.

\section*{Acknowledgments}
This research was funded by the Russian Science Foundation, grant number 22-11-00055-P.

\section*{Author declarations}

\textbf{Conflict of interest.} The authors have no conflicts to disclose.

\textbf{Author contributions.} Andrei Velichko: conceptualization, methodology, formal analysis, visualization, writing---original draft, supervision, project administration, funding acquisition, and correspondence. N'Gbo N'Gbo: software, investigation, validation, data curation, writing---review and editing. Viet-Thanh Pham: validation, scientific discussion, writing---review and editing.

\end{document}